\documentclass{pasa}%

\title[Pulsar Observations with a Phased Array Feed]{Observing Pulsars with a Phased Array Feed at the Parkes Telescope}
\author[Deng et al.]{X. Deng$^{1,2}$, A.~P. Chippendale$^{1,4}$, G. Hobbs$^1$, S. Johnston$^1$, S. Dai$^1$, D. George$^1$, M. Kramer$^{2,3}$, R. Karuppusamy$^2$, M. Malenta$^3$, L. Spitler$^2$, T. Tzioumis$^1$ \and G. Wieching$^2$\\
\affil{$^1$CSIRO Astronomy and Space Science, Australia Telescope National Facility, P.O.~Box~76, Epping NSW~1710, Australia}%
\affil{$^2$Max-Planck-Institut f\"{u}r Radioastronomie, Auf dem H\"{u}gel 69, 53121 Bonn, Germany}
\affil{$^3$Jodrell Bank Centre for Astrophysics, The University of Manchester, Alan Turing Building, Manchester M13 9PL, UK}
\affil{$^4$Email: \href{mailto:Aaron.Chippendale@csiro.au}{Aaron.Chippendale@csiro.au}}
}%

\jid{PASA}
\doi{10.1017/pas.\the\year.xxx}
\jyear{\the\year}

\usepackage[authoryear]{natbib}
\bibpunct{(}{)}{;}{a}{}{,}
\usepackage{aas_macros}
\usepackage{enumerate}
\usepackage{multirow}
\PassOptionsToPackage{hyphens}{url}\usepackage{hyperref}
\hypersetup{colorlinks,citecolor=blue,linkcolor=blue,urlcolor=blue,breaklinks=true}

\DeclareMathOperator*{\argmin}{\arg\min}
\usepackage[binary-units=true]{siunitx}
\DeclareSIUnit\jansky{Jy}
\DeclareSIUnit\parsec{pc}

\usepackage{tikz}
\usepackage{textcomp}
\usepackage{hyperref}
\usepackage{lipsum}

\begin{document}%

\begin{abstract}
  During February 2016, CSIRO Astronomy and Space Science and the Max-Planck-Institute for Radio Astronomy installed, commissioned and carried out science observations with a phased array feed (PAF) receiver system on the \SI{64}{\metre} diameter Parkes radio telescope. 
Here we demonstrate that the PAF can be used for pulsar observations and we highlight some unique capabilities. We demonstrate that the pulse profiles obtained using the PAF can be calibrated 
  and that multiple pulsars can be simultaneously observed. Significantly, we find that an intrinsic polarisation leakage of \SI{-31}{\decibel} can be achieved with a PAF beam offset from the centre of the field of view. We discuss the possibilities for using a PAF for future pulsar observations and for searching for fast radio bursts with the Parkes and Effelsberg telescopes.
\end{abstract}

\maketitle%
%
\section{INTRODUCTION}
\label{sec:intro}

The Parkes \SI{64}{\metre} diameter telescope has been used for pulsar research since the discovery of pulsars.  As the receivers and backend instruments have continued to be upgraded the telescope still produces high quality pulsar observations.  The telescope has been used to discover more than half of the known pulsars.  The majority of these pulsars were found using a 13-beam, \SI{20}{\centi\metre} multibeam receiver \citep{staveley1996parkes}. Observations of known pulsars are usually carried out with the central feed of the multibeam receiver, or with a dual-band, single-pixel receiver.   A new ultra-wide-band single-pixel receiver is currently being designed that will provide continuous coverage from \SIrange{0.7}{4}{\giga\hertz} and is expected to be installed on the telescope during 2017 \citep{Dunning2015,ManchesterPPTA2013}.

The multibeam receiver has been used extensively for pulsar surveys over the last decade \citep{keith2010high,ng2015high,lorimer2015parkes} and has been the main instrument for detecting fast radio bursts (FRBs) \citep{thornton2013population,champion2016five} to date. The multibeam receiver is sensitive, but the current FRB searches are limited by the field of view (FoV) being observed. In February 2016 a phased array feed (PAF) receiver was installed at Parkes that more than doubled the number of simultaneous beams on the sky compared with the multibeam system. This was the first long-term installation of a PAF on a high-gain, single-dish telescope with significant direct access by astronomers. The Commonwealth Scientific and Industrial Research Organisation (CSIRO) and  the Max-Planck-Institute for Radio Astronomy (MPIfR) used this unique opportunity to collaborate on wide FoV projects relating to pulsars and transient sources with this novel receiver system. We report on our results relating to studies of known pulsars in this paper. The PAF was located in the focus cabin of the Parkes telescope from February to October 2016. It was then removed for shipment to Germany where it will be installed in early 2017.


A PAF is a dense array of antenna elements at the focus of a reflector telescope and the output of these elements can be combined to form beams on the sky. The direction of these beams is controlled by varying the weighting of individual elements of the PAF.  The PAF described here \citep{hampson2012askap, Hay2008} was designed for the Australian Square Kilometre Array Pathfinder (ASKAP) telescope \citep{DeBoer2009,SchinckelASKAP2016}, but slightly modified for use on the Parkes and Effelsberg telescopes \citep{chippendale2016testing}. At present, the PAFs on ASKAP are used in standard synthesis imaging mode, with correlations between corresponding PAF beams on different antennas produced every \SI{10}{\second}. This is sufficient to see slow transients \citep{hobbs2016pilot} but not sufficient to observe pulsars with high time resolution.
 
Pulsars have been observed in a traditional pulsar observing mode with PAFs on the Westerbork Synthesis Radio Telescope. Initial profiles for two simultaneously-observed pulsars were given by \cite{van2011experimental}.  An earlier generation PAF has also been installed on a \SI{12}{\metre} diameter test-bed antenna at Parkes.  This system is currently being used to monitor PSR J0835$-$4510 (the Vela pulsar) and the results will be published elsewhere~\citep{Sarkissian2017}.

In this paper, we (1) demonstrate that a PAF on a high-gain single dish telescope can be used for pulsar observations, (2) show that such observations can be calibrated, (3) show that three pulsars can be simultaneously observed for extended periods without rotating the receiver by updating the beam positions in real time and (4) discuss the expected future use of PAFs for pulsar and transient astronomy.

\section{THE PAF INSTALLED AT PARKES}

The CSIRO PAF system \citep{hampson2012askap} was designed for the ASKAP telescope to carry out fast astronomical surveys with a wide FoV in a frequency band between \SIlist{0.7; 1.8}{\giga\hertz}.  One of these PAFs was installed on the Parkes radio telescope to test and commission the PAF  and the corresponding backend instrumentation \citep{chippendale2016testing}, to study the use of PAFs for spectral line observations, to develop software for pulsar and transient observations, and to explore the use of a PAF for studying known pulsars.

\citet{hampson2012askap} described the PAF and its associated digital processing hardware in detail. In brief, an array of 188 connected ``chequerboard'' antenna elements \citep{Hay2008} is distributed over approximately a \SI{1.2}{\metre} diameter circle.  It is a dual-polarisation receiver and each polarisation has 94 elements. The analog signals from all elements, each of up to \SI{600}{\mega\hertz} bandwidth, are transmitted to the digital receiver via RF-over-fibre links and sampled there by 12 ``Dragonfly'' digital receivers \citep{BrownDragonfly}.  The digital receivers also channelize the data to \SI{1}{\mega\hertz} via a multi-stage oversampled filterbank \citep{TuthillFilterbank}.  With 16 ports per receiver this results in a 192 port digital system, with four spare ports beyond the 188 connected to the PAF.  The digitised signals are processed by eight ``Redback'' beamformers \citep{HampsonRedback} to form up to 36 dual-polarisation beams of \SI{384}{\mega\hertz} bandwidth (\SI{48}{\mega\hertz} per beamformer) in \SI{1}{\mega\hertz} frequency channels.

\begin{figure}[htbp]
  \centering
  \includegraphics[width=0.45\textwidth]{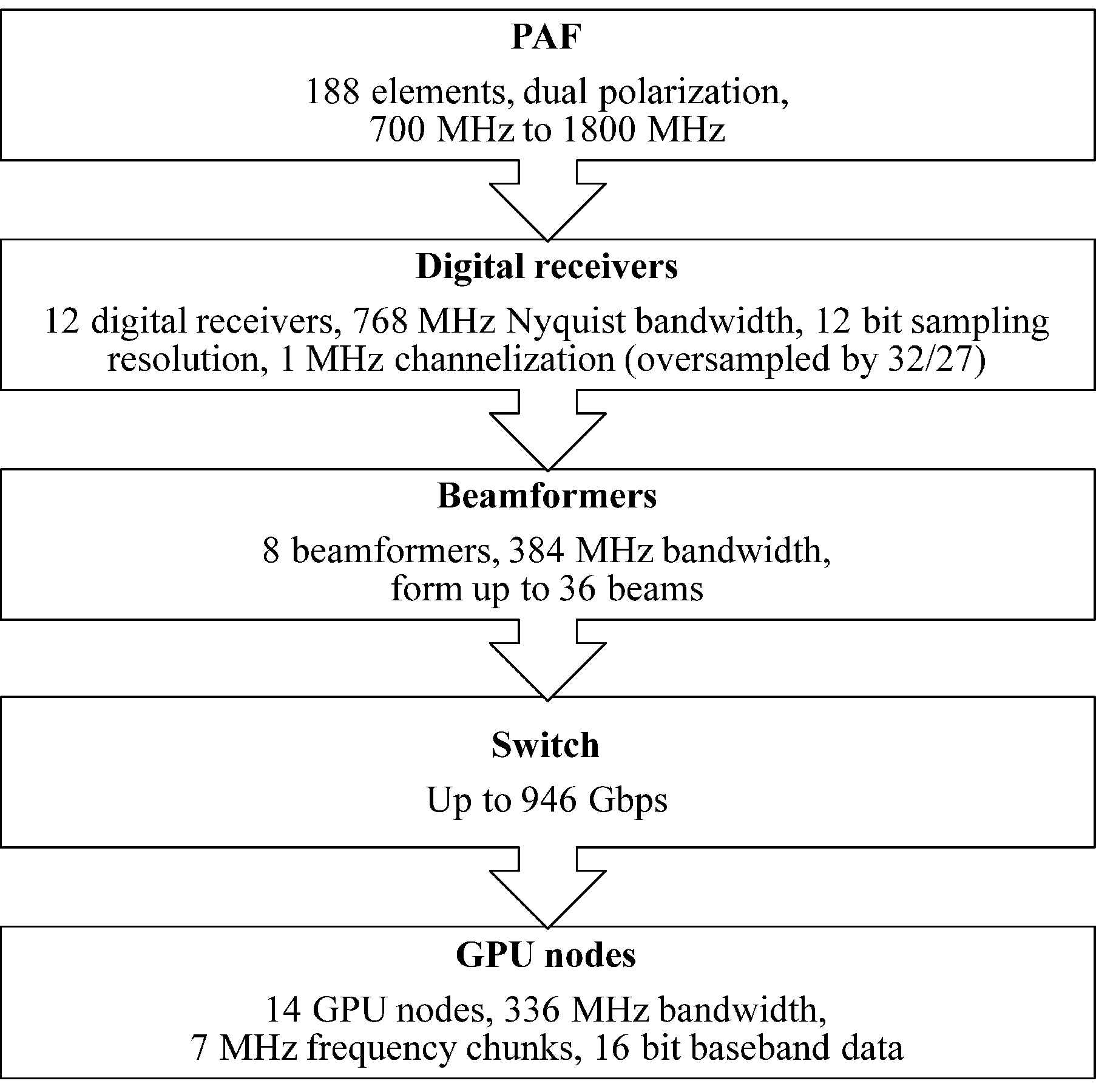}
  \caption{Schematic showing the data flow for the PAF installed at the Parkes
  radio telescope.}
  \label{fig:paf-diagram}
\end{figure}

For ASKAP, the outputs of the beamformers from different antennas are correlated at a central site for calibration and synthesis imaging.  At Parkes we stream \SI{336}{\mega\hertz} of the beamformed data (\SI{42}{\mega\hertz} per beamformer) at the full sampling rate into Graphics Processing Unit (GPU) nodes via Ethernet switches in \SI{7}{\mega\hertz} frequency chunks.  The beamformed data are scaled from \SI{16}{\bit} down to \SI{8}{\bit} on these GPU nodes. We can then (1) record \SI{8}{\bit} baseband data on the GPU nodes and fold it offline at the period of a known pulsar with the \textsc{dspsr} software package \citep{vanSratenDSPSR} or (2) fold the streaming data in ring buffers using \textsc{dspsr} in real-time. Figure~\ref{fig:paf-diagram} summarises the data flow.

As there is more radio-frequency interference (RFI) at Parkes and Effelsberg than at the ASKAP site, we used narrower bandpass filters in the PAF to reject interference from mobile phones and lower-frequency digital television services. The PAF system used at Parkes was optimised to cover the quieter \SIrange{1.2}{1.74}{\giga\hertz} band with two frequency bands covering \SIrange{1.2}{1.48}{\giga\hertz} and \SIrange{1.34}{1.74}{\giga\hertz} \citep{chippendale2016testing}.  In these modified bands the system was able to operate at Parkes with the same attenuator settings and analog-to-digital converter input levels used for ASKAP at the radio-quiet Murchison Radio-astronomy Observatory (MRO).  A further two observing bands are available with unmodified ASKAP filters covering \SIrange{0.6}{0.7}{\giga\hertz} and \SIrange{0.7}{1.2}{\giga\hertz}.  The RFI in these unmodified bands required at least \SI{10}{\decibel} more attenuation than at the MRO and varied significantly with the orientation of the \SI{64}{\metre} antenna and PAF relative to local transmitters.  In this paper, we only discuss observations in the \SIrange{1.2}{1.48}{\giga\hertz} band, which is relatively quiet with the exception of satellite navigation signals that come and go as satellites overfly the observatory.  

\subsection{ Dedispersing oversampled time series}
In this work we applied offline, incoherent dedispersion via \textsc{psrchive} \citep{hvm04} on data recorded from the oversampled polyphase filterbank designed for ASKAP \citep{TuthillFilterbank}.  The incoming data from each \SI{1}{\mega\hertz} sub-band of the filterbank are folded at the period of the pulsar and time averaged online by the \textsc{dspsr} package as described above.

Aliased signals from adjacent filterbank sub-bands distort integrated full-band pulse profiles by adding attenuated and time-shifted copies of the correct pulse profile \citep{vanSratenDSPSR}.  The attenuation is approximately the mean level of adjacent-channel leakage and the time-shift is approximately the dispersion delay between adjacent channels.  Figure 4 in \cite{vanSratenDSPSR} shows a \SI{-12}{\decibel} feature at a time-shift of \SI{100}{\micro\second} for a critically sampled FFT filterbank\footnote{Referred to as a ``deprecated'' filterbank by \cite{vanSratenDSPSR}.} with \SI{500}{\kilo\hertz} channel-width and a dispersion delay between channels of \SI{88}{\micro\second} to \SI{154}{\micro\second} over the observing band for a pulsar with $DM=71$~\SI{}{\per\centi\meter\cubed\parsec}.  The \SI{-12}{\decibel} level of the time-shifted copy in \citep{vanSratenDSPSR} is very close to the value of \SI{-12.3}{\decibel} given by averaging the spectral leakage calculated for an FFT with rectangular window over one adjacent channel. 

ASKAP's oversampled coarse filterbank provides a flatter overall passband response and reduces aliasing in its \SI{1}{\mega\hertz} sub-bands \citep{TuthillFilterbank, Tuthill2015}.  Calculating the average spectral leakage over one adjacent \SI{1}{\mega\hertz} sub-band of the ASKAP filterbank suggests we would expect a time-shifted copy of the pulse profile at the level of \SI{-27}{\decibel} in the averaged pulse profiles presented in this paper.  For an observation of pulsar with $DM=71$~\SI{}{\per\centi\meter\cubed\parsec} in the \SIrange{1200}{1480}{\mega\hertz} MPIfR PAF band range we would expect this additive distortion to appear at a time-shift equal to the inter-channel dispersion delay of \SIrange{182}{341}{\micro\second}.

In fact, near artefact-free dedispersion of pulsar measurements made with the oversampled ASKAP filterbank should be possible if the \SI{1}{\mega\hertz} channels are further channelized.  This would allow the unwanted band edges of the oversampled \SI{1}{\mega\hertz} sub-bands to be discarded before coherent dedispersion and frequency averaging of the pulse profile over the full observing band.  This can be accomplished by channelizing each \SI{1}{\mega\hertz} sub-band into $N$ fine channels then, for an oversampling ratio of $\rho = 32/27$, discarding $N(1-\rho^{-1})/2$ fine channels from both ends of each \SI{1}{\mega\hertz} sub-band before further processing.  By averaging the spectral leakage from one adjacent sub-band over only the retained fine channels we expect the erroneous copy of the pulse profile to appear at \SI{-74}{\decibel}. 




\color{black}
\section{RESULTS}
Throughout the PAF commissioning period we observed PSRs~J0437$-$4715, J0742$-$2822, J0835$-$4510, J1559$-$4338 and J1644$-$4559. We also identified a non-globular-cluster sky region in which multiple pulsars, with flux density greater than \SI{0.5}{\milli\jansky}, are in the PAF FoV. 

First, we tested the entire observing system with the bright pulsars, PSRs~J0742$-$2822, J0835$-$4510, J1559$-$4338 and J1644$-$4559.  This allowed us to test the beam positions and the correct ordering of the frequency channels in the resulting data files. We then observed the bright millisecond pulsar PSR~J0437$-$4715 to study the achievable timing precision and calibration method. Finally we simultaneously observed three pulsars to test the timing quality including beam updates to track sources drifting through the FoV.

For all of this work we used maximum signal-to-noise ratio (SNR) beam weights calculated via the same method used for performance evaluation of the ASKAP Boolardy Engineering Test Array (BETA, \citealt{McConnellBETA2016}). We applied this beam weights calculation algorithm independently for each set of 94 PAF elements with common polarisation.  Thus, as with the ASKAP BETA work, we only weighted 94 elements with matching polarisation into each beam.  This makes the beam polarisations line up with the native PAF element polarisations, which are inherently linear with low and stable leakages \citep{SaultBETA2014}. This also provides a clear link between the beam polarisation and the physical orientation of the PAF, which is convenient during commissioning.  Unlike the BETA work, we used strong extragalactic sources such as Virgo A for beam weight calculation instead of the Sun which is too extended for the narrower beam of the \SI{64}{\metre} Parkes telescope.

\begin{figure}[tbp]
  \centering
 \includegraphics[width=0.4\textwidth]{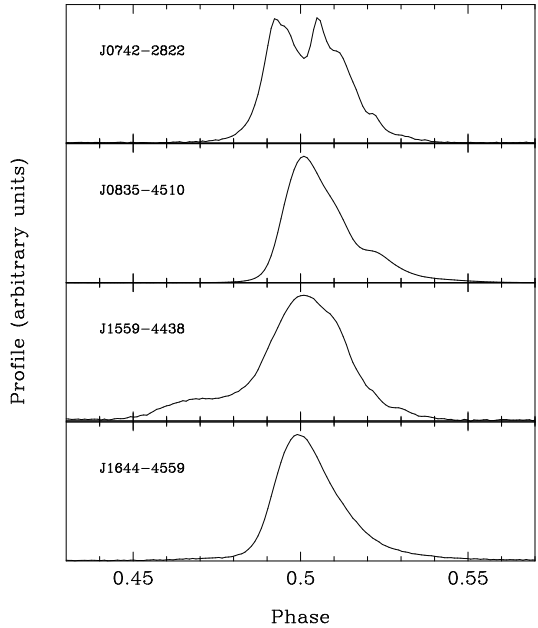}
  \caption{Total intensity (Stokes I) profiles of four pulsars we measured as a test of the entire PAF observing system.  The integration times of these observations are \SI{1}{\hour}, \SI{10}{\minute}, \SI{40}{\minute} and \SI{10}{\minute} respectively. We centred these profiles and zoomed them to the pulse-phase range from 0.43 to 0.57.}
  \label{fig:profiles}
\end{figure}

 
Figure~\ref{fig:profiles} shows the uncalibrated total intensity profiles for PSRs J0742$-$2822, J0835$-$4510, J1559$-$4438 and J1644$-$4559. We compared them with the profiles observed with Parkes project P574 (see \citealt{weltevrede2008profile}) in January 2014 and the maximum relative deviation is less then \SI{10}{\percent} for the latter three of the four pulsars\footnote{To assess pulse profile similarity in phase with \textsc{psrchive} for each pulsar, we normalised the aligned profiles and got the deviation between them. We repeated that for the latter three pulsars and the maximum relative deviation is the maximum deviation given by the whole procedure.}.
 
Our observed profile for PSR~J0742$-$2822 is slightly different from that given in the archive. This pulsar has a profile, which is variable on short timescales (see \citealt{keith2013connection}) and observations taken with a traditional receiver only one day after the observation obtained with the PAF show the same structure in the profile. We therefore conclude that the PAF functions well in Stokes I and will discuss polarisation calibration in the next section. 
 
 


%
%

\begin{figure}[tbp]
  \centering
  \includegraphics[width=0.4\textwidth,angle=-90]{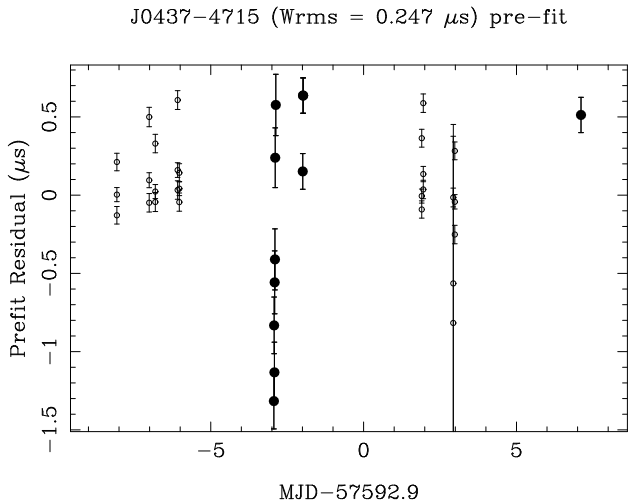}  
  \caption{Timing residuals of PSR J0437$-$4715. The open and solid circle symbols are for the timing residuals obtained with a traditional receiver and the PAF system respectively.}
  \label{fig:0437-residuals}
\end{figure}

In order to demonstrate that pulse arrival times can be measured over days we made multiple observations of the brightest millisecond pulsar, PSR~J0437$-$4715.  This pulsar can be timed with an rms timing precision of \SI{100}{\nano\second} over years using existing receiver systems at the Parkes telescope (e.g. \citealt{manchester2013parkes}).  We used the PAF to make multiple observations of this pulsar on 21 July 2016 and then follow-up observations on 22 July 2016 and 31 July 2016.  In Figure~\ref{fig:0437-residuals} we compare the timing residuals obtained using the PAF (\SI{10}{\minute} integrations, solid circle symbols) with those being obtained for the Parkes Pulsar Timing Array (PPTA) project using the traditional instrumentation (\SI{1}{\hour} integrations, open circle symbols, see \citealt{manchester2013parkes} for detail).  We note that the timing residuals from the PAF are consistent with those from the PPTA data. 

We did not apply any polarisation calibration to the PAF data in Figure \ref{fig:0437-residuals} and therefore expect to see increased scatter.  The scatter in the PAF timing is approximately \SI{2}{\micro\second} but we show in the next section that polarisation calibration can reduce this by more than a factor of two.  For comparison, the jitter noise of PSR J0437$-$4715 with \SI{10}{\minute} integration time is approximately \SI{100}{\nano\second}~\citep{shannon2014limitations} and the PPTA time of arrival (ToA) uncertainties are approximately \SI{100}{\nano\second}. 

A simple calculation based on the radiometer equation suggests that the PAF ToA uncertainties, shown by the error bars in Figure \ref{fig:0437-residuals}, should be $\sim$5.5 times larger due to the increased system temperature and slightly wider bandwidth of the PAF compared to the multibeam receiver.  This is consistent with the uncertainties in our observed arrival times.

\subsection{Calibrating the pulsar data}
\label{sect:calibration}
Pulsars are often highly polarised.  In order to form stable pulse profiles for high-precision timing, to study the emission mechanism or to analyse the scintillation properties of a pulsar, it is necessary to carry out polarisation and/or flux calibration. The polarisation properties of a signal are fully described using the four Stokes parameters: I, Q, U and V.  The measured Stokes parameters will have been affected by the observing instrument and telescope.  Calibrating traditional, single-pixel receivers is non-trivial and often relies on the injection of a pulsed-calibration signal (e.g. \citealt{van2004radio}). The PAF does not include such a noise calibration system.  Instead a broad-band radiator was installed on the telescope surface and this has been used to study methods for forming the PAF beams and making them stable with time.

In the work described here we solve for the instrumental polarisation parameters of a PAF beam that minimise the mean square difference between new PAF measurements of the polarisation profile of PSR~J0437$-$4715 and a corresponding archived reference measurement that is known to be accurately calibrated.   Our approach is very similar to that of \cite{JohnstonCalibration} and we use a reference measurement from \cite{ShiPolProfiles} that was made using the central feed of the \SI{20}{\centi\metre} multibeam receiver.  The multibeam receiver is carefully calibrated and in regular use for precise measurements of pulsars.  


The relationship between the Stokes parameters of the incident wave and beamformed voltages ($\mathbf{S}_i$ and $\mathbf{S}_o$ respectively) can be modelled as \citep{warnick2012polarimetry}
\begin{equation}
  \label{eq:S-relationship}
  \mathbf{S}_i=\mathbf{M}^{-1}\mathbf{S}_o
\end{equation}
where $\mathbf{S}=[I, Q, U, V]^{T}$ is a column vector of Stokes parameters, $\mathbf{M}=\mathbf{T}\left(\mathbf{J}\otimes\mathbf{J}^*\right)\mathbf{T}^{-1}$ is the Mueller matrix of the PAF system and $\mathbf{T}$ is the transformation matrix between the Stokes parameters and the coherency vector 
\begin{equation}
  \label{eq:tran_s_c}
  \mathbf{T}=
      \left[\begin{array}{cccc}  
              1&0&0&1\\
              1&0&0&-1\\
              0&1&1&0\\
              0&-i&i&0\\
            \end{array}\right].
        \end{equation}
        $\mathbf{J}=\mathbf{J}_\text{cal}\mathbf{J}_\psi$ is the Jones matrix of the combined PAF  and \SI{64}{\metre} antenna that comprises a factor due to the rotation of the feed by the parallactic angle, $\psi$, as the altazimuth mounted antenna tracks the source
\begin{equation}
  \label{eq:jonespsi}
\mathbf{J}_\psi = \left[\begin{array}{cc}  
          \cos\psi & \sin\psi\\
          -\sin\psi & \cos\psi\\
        \end{array}\right]
\end{equation}
and a generalised Jones matrix that represents the instrumental polarisation of the PAF     
\begin{equation}
  \label{eq:jonescal}
\mathbf{J}_\text{cal} = \left[\begin{array}{cc}  
          A & B\\
          C & D\\
        \end{array}\right].
    \end{equation}
Here $B$, $C$ and $D$ are complex, but we force $A$ to be real to account for the fact that this technique cannot calibrate a phase or delay difference between the PAF and the receiver used to make the archived measurement.
$\mathbf{J}^*$ is the complex conjugate of $\mathbf{J}$ and $\otimes$ is the Kronecker product operator.

The calibration problem can be expressed as 
\begin{equation}
  \label{eq:cal-problem}
  \argmin_{\mathbf{J}_\text{cal}} \left<|\mathbf{S}_i-\mathbf{S}_\text{ref}|^2\right>
\end{equation}
where $\mathbf{S}_\text{ref}$ are the ``true'' Stokes parameters provided by the archived measurement and $\left<\cdot\right>$ denotes averaging over pulse-phase bins and PAF observation epochs.  This equation can be uniquely solved if we include measured profiles of the pulsar from multiple observation epochs with different parallactic angles. 

\begin{figure*}[htbp]
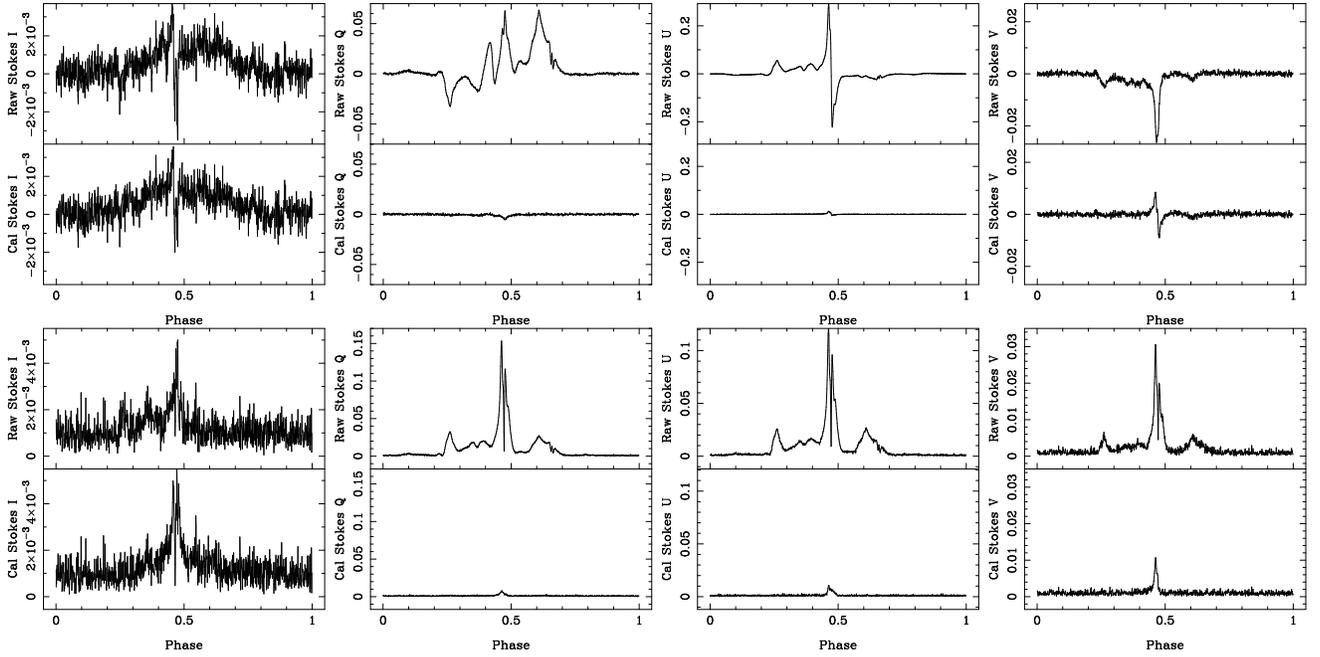

  \centering

  \includegraphics[width=0.245\textwidth,angle=-90]{mean-I.ps}
  \includegraphics[width=0.245\textwidth,angle=-90]{mean-Q.ps}
  \includegraphics[width=0.245\textwidth,angle=-90]{mean-U.ps}
  \includegraphics[width=0.245\textwidth,angle=-90]{mean-V.ps}  
  \includegraphics[width=0.245\textwidth,angle=-90]{std-I-new.ps}
  \includegraphics[width=0.245\textwidth,angle=-90]{std-Q-new.ps}
  \includegraphics[width=0.245\textwidth,angle=-90]{std-U-new.ps}
  \includegraphics[width=0.245\textwidth,angle=-90]{std-V-new.ps}
  \caption{Error in Stokes polarisation parameters of PSR J0437$-$4715 for PAF observations with respect to PPTA reference observations  (with Stokes I, Q, U and V from left to right).  The figures on the top row are for the mean of the difference (between these four observed profiles and the reference template) and the figures on the bottom row are for the standard deviation of the difference. The upper panel of each figure shows the uncalibrated result and the bottom panel of each figure represents the calibrated result.}
  
  \label{fig:stoke-0437}
\end{figure*}

We followed these steps to calibrate the data:
\begin{enumerate}
\item The weighted profile of PSR J0437$-$4715 in the \SI{20}{\centi\metre} observing band was downloaded from the PPTA pulsar profile collection \citep{ShiPolProfiles}\footnote{Available via the CSIRO Data Access Portal at \url{http://doi.org/10.4225/08/54F3990BDF3F1}.} of the Parkes Observatory Pulsar Data Archive \citep{hobbs2011parkes}. 
\item The band between \SIlist{1304; 1465}{\mega\hertz} was extracted to avoid RFI contamination from satellite navigation services at lower frequencies and to avoid exceeding the PAF observing band edge at higher frequencies. This extracted profile was used as our polarisation template\footnote{This profile was based on the multibeam receiver and calibrated using the \textsc{pcm} calibration method within the \textsc{psrchive} pulsar data processing software package \citep{hvm04}.}.
\item The Stokes parameters of this reference template were normalised by its peak Stokes I value.
\item The PAF was used to observe PSR~J0437$-$4715 four times on 6 June 2016 with each observation lasting one hour\footnote{To confirm that the calibration works for a pulsar that is not directly in the pointing direction of the telescope we used a beam with \SI{0.215}{\degree} offsets in both elevation and cross-elevation.  We tracked PSR~J0437$-$4715 with complementary offsets in elevation and cross-elevation to keep this pulsar in the offset beam.}.
\item The PAF data were binned to \SI{10}{\minute} sub-integrations and the data were summed in frequency between \SIlist{1304; 1465}{\mega\hertz}.
\item The Stokes parameters of each \SI{10}{\minute} PAF measurement were normalised to its peak Stokes I value. \label{step:Normalised}
\item The observed profiles were aligned in pulse phase with the reference template\footnote{The alignment shift was determined by cross-correlating the Stokes I profile from the PAF observation with that of the reference.  The same alignment shift was applied to all Stokes parameters of the measured profile.}. \label{step:align}
\item The minimisation problem of Equation~\eqref{eq:cal-problem} was solved to calculate $\mathbf{J}_\text{cal}$ using all pulse-phase bins for all of these \SI{10}{\minute} PAF profiles.
\item The \SI{10}{\minute} PAF profiles were calibrated using Equation~\eqref{eq:S-relationship}.
\end{enumerate}

Figure~\ref{fig:stoke-0437} shows the difference between the observed profiles\footnote{These calibrated \SI{10}{\minute} profiles were normalised (as at Step \ref{step:Normalised}) and aligned (as at Step \ref{step:align}). We formed four calibrated \SI{1}{\hour} profiles by adding calibrated \SI{10}{\minute} profiles in groups of six.} and template (with and without calibration). The uncalibrated Stokes I profile is very close to the true value and that the uncalibrated Stokes Q, U and V profiles are significantly deviant.  Post-calibration, the standard deviation of the differences between the observed profiles and the reference profile, for all Stokes parameters, is better than \SI{1}{\percent} of the peak amplitude of Stokes I.  We note that the profiles are significantly improved after calibration, but they are not perfect.  Such imperfection could arise from intrinsic pulse shape changes since the PPTA measurement, imperfect calibration or ionospheric effects. 

We tested for possible Faraday rotation in the ionosphere by repeating the calibration and simultaneously solving for an independent rotation measure, $\mathrm{RM}$, at each \SI{10}{\minute} observation interval. We forced $\mathrm{RM}=0$ at the first \SI{10}{\minute} interval so this defined a rotation of the received polarisation plane by $\mathrm{RM}\lambda^2$ at each interval with respect to the first (where $\lambda$ is the wavelength for each observation channel).  The fit suggests there was a change of $\mathrm{RM}$ of approximately \SI{1}{\radian\per\square\metre} over the first three hours of the observation.  We explored the sensibility of this fit by using \textsc{rmextract}\footnote{Available at \url{https://github.com/maaijke/RMextract}.} to calculate expected ionospheric $\mathrm{RM}$ from Global Positioning System measurements of total electron content, and the World Magnetic Model\footnote{See \url{https://www.ngdc.noaa.gov/geomag/WMM/DoDWMM.shtml}.}.  This also suggested a \SI{1}{\radian\per\square\metre} change in $\mathrm{RM}$, but over the fourth to ninth hours of the observation and so our results are not conclusive, but changes in RM of this size are not unexpected.

\begin{figure}[tbp]
  \centering
\includegraphics[width=0.33\textwidth, angle=-90]{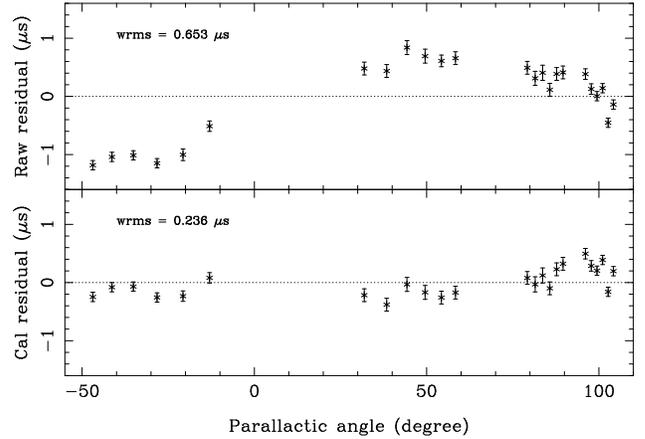}
\caption{Timing residuals of PSR J0437$-$4715. The upper and lower panels represent the timing residuals before and after polarisation calibration.\label{fig:0437-compare}}
\end{figure}

We formed an analytic, noise-free standard template for PSR~J0437$-$4715 using the \textsc{psrchive} package \textsc{paas} and determined pulse arrival times using \textsc{pat} for the uncalibrated and calibrated \SI{10}{\minute} profiles. Figure~\ref{fig:0437-compare} represents the resulting timing residuals with respect to the timing model we used for Figure~\ref{fig:0437-residuals}. The polarisation calibration reduces the weighted rms of the timing residuals from \SIrange{0.653}{0.236}{\micro\second}, an improvement of more than a factor of two.

\subsection{Simultaneous observations of multiple pulsars}
\label{sect:timing_multiple}
As noted above, the primary advantage of using a PAF, instead of a single-pixel receiver, is that the PAF provides a wide FoV. We used the Australia Telescope National Facility (ATNF) pulsar catalogue~\citep{manchesteraustralia} to select regions of the sky in which multiple pulsars can be observed in a single observation with the PAF at Parkes. We removed regions that contained globular clusters as pulsars in those clusters could be observed with a single pixel receiver and therefore using the PAF is not advantageous. We then selected a region in which three pulsars could easily be detectable with the PAF.   

\begin{figure}[tbp]
  \centering
\includegraphics[width=0.47\textwidth, angle=-90]{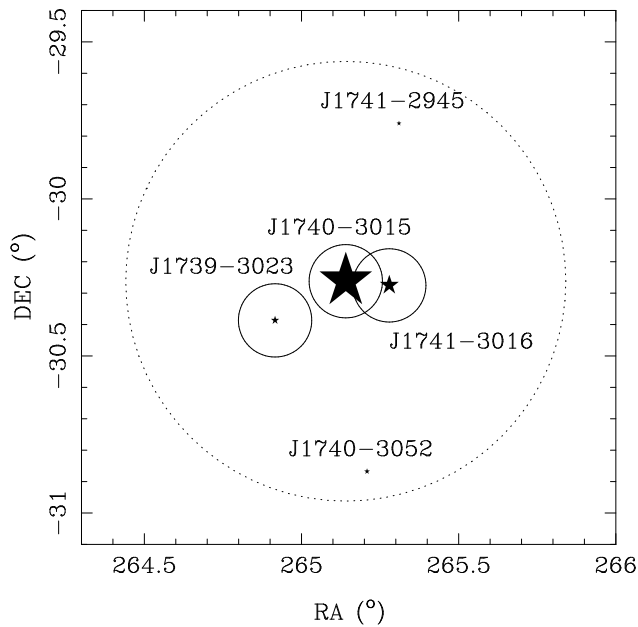} 
\caption{Region around PSR J1740$-$3015. Pulsars are indicated with star symbols and the names of these pulsars are given in the figure. The size of each star symbol indicates the flux density of that pulsar (the larger the symbol, the higher the flux density). Pulsars surrounded by \SI{7}{\arcminute} radius solid circles were observed simultaneously. The dotted circle gives an approximate indication of the sky region that could be observed using all 36 beams and shows that the PAF could observe 5 pulsars simultaneously in this field with appropriate backend configuration.\label{fig:pulsar-region}}
\end{figure}

The pulsars in this region are listed in Table~\ref{tab:=region} and their positions in the FoV are indicated in Figure~\ref{fig:pulsar-region}. For each pulsar the table contains the pulsar name, its right ascension and declination, dispersion measure (DM), the radial distance (in degrees) from the central pulsar position and the flux density in the \SI{20}{\centi\metre} observing band.  Figure~\ref{fig:pulsar-region} shows that we could have observed five pulsars simultaneously in the PAF FoV, but we were restricted to three due to the limited backend configurations available during commissioning.  We were also limited to observe with \SI{112}{\mega\hertz} bandwidth for each of the three pulsars but should be able to achieve \SI{336}{\mega\hertz} bandwidth per pulsar in the future.

\begin{table*}[tbp]
  \centering
  \caption{Simultaneously observed pulsars with the PAF system.\label{tab:=region}}
    \begin{footnotesize}
\begin{tabular}{llllll}
  \hline
PSR Name      &RAJ2000&DECJ2000&DM&RD&S1400\\
              &(hh:mm:ss.s)&(+dd:mm:ss)&(\si{\cm}$^{-3}$pc)&(\si{\degree})&(\si{\milli\jansky})\\
 
  \hline
J1740$-$3015    &17:40:33.8  &$-$30:15:43     &152.15 &0.00      &6.40  \\
J1741$-$3016    &17:41:07.0  &$-$30:16:31     &382.00   &0.12      &2.30  \\
J1739$-$3023    &17:39:39.8  &$-$30:23:12     &170.00   &0.23      &1.00  \\
\hline
\end{tabular}
\end{footnotesize}
\end{table*}

\begin{figure}[tbp]
  \centering
  \includegraphics[width=0.47\textwidth, angle=-90]{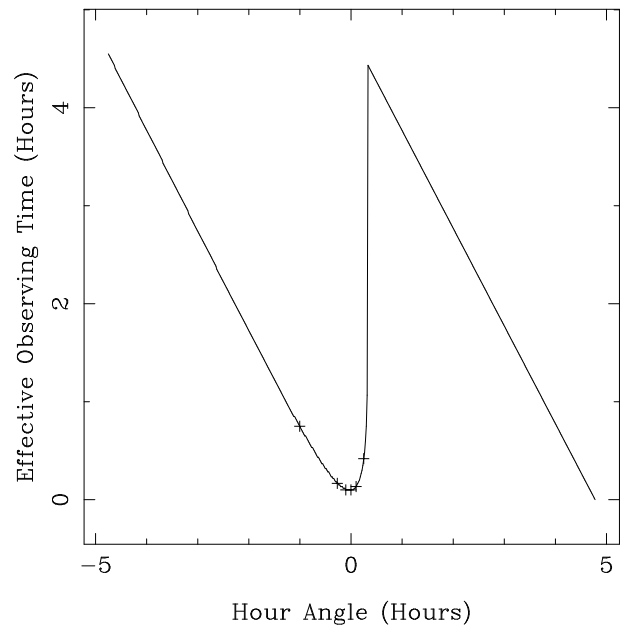} 
  \caption{Effective observing time of PSR J1741$-$3016 with an outer beam when the central beam is tracking PSR J1739$-$3023. We commence observing PSR J1739$-$3023 with the peak of an outer beam and record the end of effective observing time when the pulsar crosses the half-power point of that beam or sets below the elevation limit of the telescope.  We calculate the effective observing time for different hour angles from rise to set at one-minute intervals. Plus symbols in the figure indicate the hour angles at which we uploaded beam weights.}
  \label{fig:time-hour-angle}
\end{figure}

On 22 September 2016 from UTC 06:41, we observed the three pulsars simultaneously. We tracked PSR~J1740$-$3015 and so this pulsar remained in the centre of the FoV throughout the observation. However, as the Parkes telescope has an altazimuth mount and we are unable to rotate the PAF, the positions of the other two pulsars in the FoV changed throughout the observation.  They will drift out of non-central beams if these beams are fixed. Figure~\ref{fig:time-hour-angle} shows the effective observing time of PSR J1739$-$3023 in an outer beam as a function of the hour angle. The pulsar drifts quickly out of the beam near transit (and also the observing duration becomes small when the source is close to setting). 

We formed beams pointing at PSRs~J1739$-$3023 and J1741$-$3016 at the beginning and updated the beam positions by uploading new beam weights when the pulsar would have drifted through the beam in order to keep tracking each source (the hour angles at which we uploaded beam weights are shown with plus symbols in Figure~\ref{fig:time-hour-angle}).  We folded each \SI{10}{\second} of data for each pulsar online to form a folded-profile. We finally averaged six adjacent profiles together to form a high $S/N$ profile every minute.  Our observation lasted approximately \SI{1.5}{\hour}.  

\begin{figure}[tbp]
  \centering
  \includegraphics[width=0.47\textwidth]{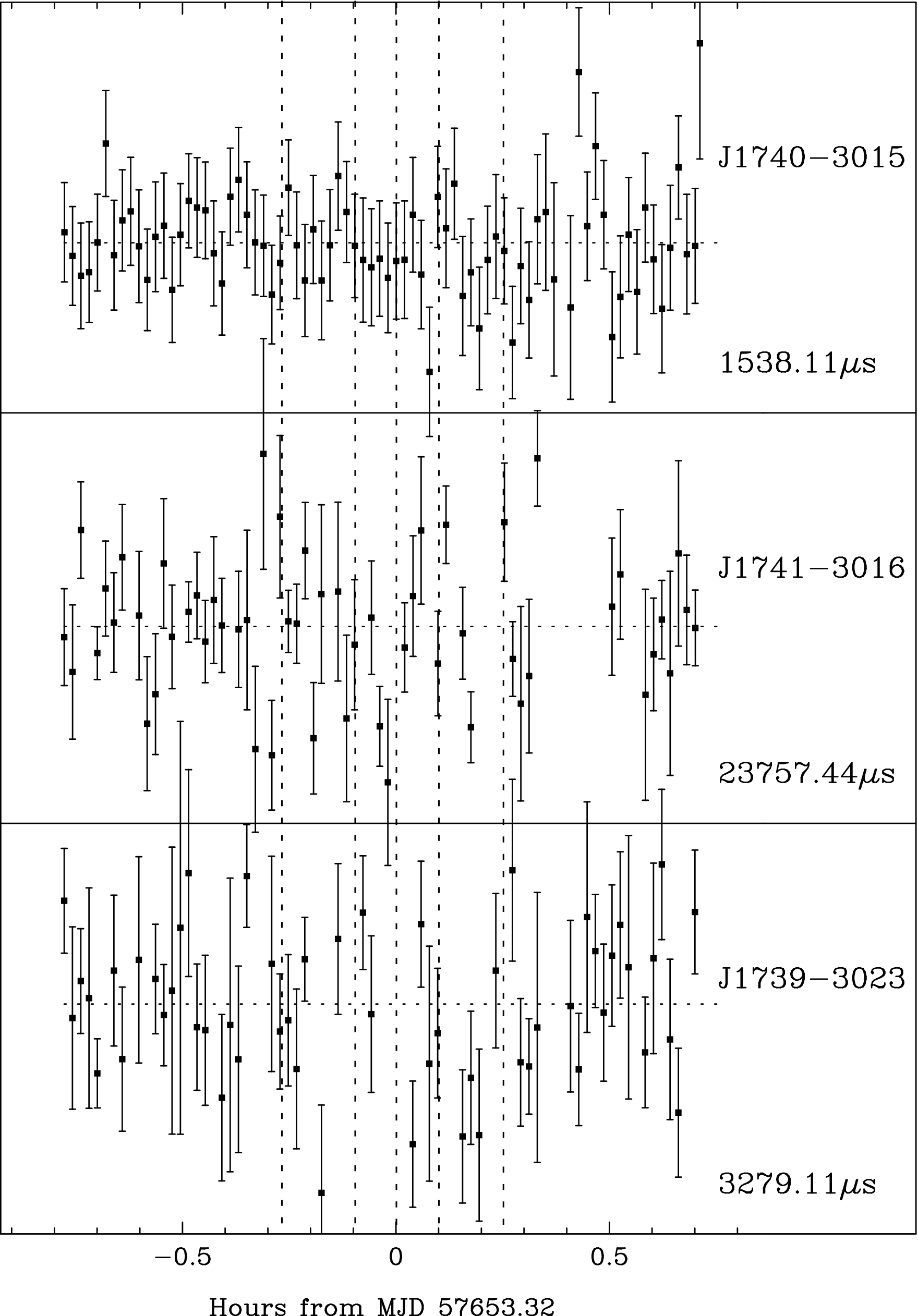} 
  
  \caption{Long-duration simultaneous pulsar observation with beam tracking. Timing residuals for different pulsars are shown in different panels and the range from the minimum to the maximum residual are \SIlist{1.5; 23.8; 3.3}{\milli\second} respectively. Vertical dashed lines indicate when beamformer weights were updated.}
  \label{fig:multi-with-track}
\end{figure}

We formed an analytic, noise-free standard template for each pulsar using the \textsc{psrchive} package \textsc{paas} and determined pulse arrival times using \textsc{pat}.  Figure~\ref{fig:multi-with-track} shows the resulting timing residuals with respect to the timing model for each pulsar in the ATNF pulsar catalogue \citep{manchesteraustralia}, obtained using the timing package \textsc{tempo2} \citep{G-Hobbs_2006, R-Edwards_2006}. The vertical lines in Figure~\ref{fig:multi-with-track} indicate the times we updated beam positions.  We clearly see that we can simultaneously observe three pulsars and their ToA uncertainties do not significantly change for the pulsars in the outer-beams throughout the observation.   


\section{DISCUSSION}

\subsection{Timing Pulsars Precisely with PAFs}

Although we are far from recommending that the current generation of PAFs be used for high precision pulsar timing, we note that the PAF design has no intrinsic hardware or other technical issues that would prohibit high precision timing.  For instance, some of the most sensitive millisecond pulsar timing data sets are currently being achieved by the Parkes Pulsar Timing Array (PPTA) project (see, e.g., \citealt{manchester2013parkes} for an overview and \citealt{shannon2014limitations} for an application of such data sets to gravitational wave searches).  The most precise data sets in the PPTA are obtained with an incoherent dedispersion system similar to that used with the PAF in this paper.  The various backend instruments have timing offsets that can vary during upgrades, but it is now possible to measure such time offsets either by comparing different backend receivers or by observing a known stable source.  

The most commonly-used PPTA receiver is the central-beam of the Parkes 13-beam receiver \citep{staveley1996parkes}. However, it was not specifically designed for high-precision pulsar timing. In particular the polarisation purity is poor and \citet{van2004radio} clearly identifies significant cross-coupling effects within the receiver.  None of this affects the achievable timing precision assuming that the data are properly calibrated. We therefore highlight that high precision pulsar timing does not require excellent polarisation purity before calibration.

The polarisation purity of \SI{-40}{\decibel} required for precision timing as reported by \cite{cordes2004pulsars} refers to a post-calibration result and not the intrinsic purity of the receiver (see also \citealt{foster2015intrinsic}).   Such calibration is carried out for the PPTA project using observations of PSR~J0437$-$4715 from rise-to-set (e.g., \citealt{van2004radio}). The calibration observations described in Section \ref{sect:calibration} are restricted in number and parallactic-angle coverage, yet we were able to use a single Jones matrix $\mathbf{J}_\text{cal}$ to describe the PAF instrumental polarisation over a \SI{161}{\mega\hertz} band and four \SI{1}{\hour} observations spread over \SI{6}{\hour}.  This indicates good instrumental stability. Analysing the resulting Jones matrix $\mathbf{J}_\text{cal}$ shows that the \textit{intrinsic polarisation leakage}\footnote{\cite{foster2015intrinsic} defines intrinsic polarisation leakage as ${-10\log_{10}(\text{IXR})}$ where IXR is the intrinsic cross-polarisation ratio defined by \cite{carozzi2011fundamental}.  These metrics are defined independently of the coordinate system and relate closely to polarimetric errors after full calibration.}, for a PAF beam offset by \SI{0.215}{\degree} in both elevation and cross-elevation, is \SI{-31}{\decibel}.  We note that \cite{foster2015intrinsic} found that only a limited fractional improvement in pulsar timing capabilities is achieved by pushing a feed's intrinsic polarisation leakage below \SI{-30}{\decibel}.



In summary, we have successfully timed a millisecond pulsar despite the relatively high system temperature of the current PAF.  With sufficient observations, all the techniques used for the PPTA could be applied to improve the precision of the PAF observations.



\subsection{Future Opportunities}
\subsubsection{Speeding up Pulsar Monitoring}
Current long-term monitoring of pulsars at Parkes uses a single pixel receiver system with a \SI{14}{\arcminute} full-width-at-half-maximum beamwidth.  Here we calculate the reduction in pointings that would be obtained if a PAF was used for these observations.  We first consider all the pulsars that can be observed by Parkes (i.e. declination $<$\SI{20}{\degree}) and are not in globular clusters.  The total number of pulsars and the number of pointings (to cover all of these pulsars if we observe with a PAF) are listed in Table~\ref{tab:=multi}. We then repeat this calculation for the Effelsberg telescope and assume a declination limit of $>$\SI{-30}{\degree}.  Finally, we repeat the calculation for the 321 pulsars that are currently being observed for the young-pulsar timing project at Parkes (P574, \citealt{kerr2014young}). We can see from Table~\ref{tab:=multi} that a PAF could reduce pointings by about \SI{32}{\percent} at Parkes, by \SI{28}{\percent} at Effelsberg and by \SI{19}{\percent} for the on-going young pulsar timing project at Parkes. Further work is required to translate the reduction in the number of pointings into a saving in observing time given the difference between the system temperature and bandwidths of the various observing systems. However, for the P574 project, the observing time is dictated by the time taken for the pulsar profile to become stable (i.e., several thousand rotations) rather than to reach a certain $S/N$ threshold. Hence for this project, the reduction in the number of pointings translates directly to an overall saving in observing time.




\begin{table}[tbp]
  \centering
  \caption{Efficiency of pulsar observation with PAF.\label{tab:=multi}}
    \begin{footnotesize}
\begin{tabular}{lccc}
  \hline
  Telescope/Project  &Pulsars   & PAF      &Reduction\\
                     &          &Pointings &(\si{\%})\\
\hline
  Parkes     &2018	&1365	&32.4\\
  Effelsberg &1404	&1011	&28.0\\
  P574       &321       &260    &19.0\\
\hline
\end{tabular}
\end{footnotesize}
\end{table}





Millisecond pulsars are observed for high-time precision experiments.  For instance, pulsar timing array projects (e.g. \citealt{manchester2013parkes,verbiest2016international, babak2016european}) observe these pulsars with the primary goal of detecting ultra-low-frequency gravitational waves.   Such a detection would require the unambiguous signature of a quadrupolar angular correlation in the pulsar timing residuals.  \cite{tiburzi2016study} demonstrated that without sufficient sky coverage it is possible to falsely detect a gravitational wave background signal.  They showed that it was necessary to observe pulsars that were close together (as well as pulsars separated far apart).  

For the Parkes Pulsar Timing Array project, the closest pulsar pair is  PSR J2129$-$5721 and PSR J2241$-$5236.   Even with the PAF these pulsars are too widely spaced for simultaneous observations.  However, for the International Pulsar Timing Array (IPTA), two pulsars (PSRs~J1910$+$1256 and J1911$+$1347) are separated by only \SI{0.8}{\degree} \citep{verbiest2016international}.  Not only would the use of a PAF to observe these pulsars reduce the required observation time, it would also enable simultaneous measurements of instrumental offsets with two precisely-timed pulsars.  

\subsubsection{Detecting and Localising FRBs}
One of the main science goals for the PAF on Effelsberg is the discovery of FRBs. Currently, Effelsberg is equipped with a 7-beam multibeam receiver and consequently the PAF with its 36 beams yields a five-fold increase in sky coverage for FRB searches. Current FRB results \citep{ravi2016magnetic} seem to indicate that sky area is more important than sensitivity for FRB detection. 

In addition to the increased sky coverage, the PAF fully samples the focal-region field unlike the multibeam which is highly undersampled. This implies that the PAF can measure FRB positions with sub-arcminute accuracy, unlike the multibeam that yields positions with errors greater than ten arcminutes. Determining accurate positions is crucial for understanding the nature of FRBs (e.g. \citealt{keane2016fast}).

\section{CONCLUSION}

We have demonstrated that pulsars can be observed using a PAF on a high-gain, single-dish telescope.  We can produce calibrated profiles, form timing residuals without any unexpected instrumental offsets, and observe multiple pulsars simultaneously whilst accounting for pulsars drifting through the FoV. 

\section{ACKNOWLEDGEMENTS}
The Parkes radio telescope is part of the ATNF, which is funded by the Commonwealth of Australia for operation as a National Facility managed by the CSIRO. The MPIfR financed the PAF discussed in this paper and its modification to cope with a less radio-quiet site. Installation and operation of the PAF at Parkes has benefited from the contributions of R. J. Beresford, M. Leach, M. Marquarding, S. Broadhurst,  D. Craig, J. Crocker, R. Kaletsch, B. Preisig, K. Reeves, Dr. J. E. Reynolds, T. Ruckley, M. Smith, E. Troup and Dr. J. Tuthill.  Exploration of ionospheric Faraday rotation was suggested by Dr. S. Ord and supported by Dr. E. Lenc. The calibration work benefits from discussions with Dr. K.~J. Lee.

\bibliographystyle{pasa-mnras}
\bibliography{IEEEabrv,journals,eng,psrrefs,modrefs,chi139}

\begin{thebibliography}{}
\makeatletter
\relax
\def\mn@urlcharsother{\let\do\@makeother \do\$\do\&\do\#\do\^\do\_\do\%\do\~}
\definecolor{darkblue}{rgb}{0,0,0.597656}
\def\mndoi{\begingroup\mn@urlcharsother \@ifnextchar [ {\mndoi@} {\mndoi@[]}}
\def\mndoi@[#1]#2{\def\@tempa{#1}\ifx\@tempa\@empty \href
  {http://dx.doi.org/#2} {\textcolor{darkblue}{doi:#2}}\else \href
  {http://dx.doi.org/#2} {\textcolor{darkblue}{#1}}\fi \endgroup}
\def\mn@eprint#1#2{\mn@eprint@#1:#2::\@nil}
\def\mn@eprint@arXiv#1{\href {http://arxiv.org/abs/#1} {{\tt arXiv:#1}}}
\def\mn@eprint@dblp#1{\href {http://dblp.uni-trier.de/rec/bibtex/#1.xml}
  {dblp:#1}}
\def\mn@eprint@#1:#2:#3:#4\@nil{\def\@tempa {#1}\def\@tempb {#2}\def\@tempc
  {#3}\ifx \@tempc \@empty \let \@tempc \@tempb \let \@tempb \@tempa \fi \ifx
  \@tempb \@empty \def\@tempb {arXiv}\fi \@ifundefined
  {mn@eprint@\@tempb}{\@tempb:\@tempc}{\expandafter \expandafter \csname
  mn@eprint@\@tempb\endcsname \expandafter{\@tempc}}}

\bibitem[\protect\citeauthoryear{{Babak} et~al.,}{{Babak}
  et~al.}{2016}]{babak2016european}
{Babak} S.,  et~al., 2016, \mndoi [\mnras] {10.1093/mnras/stv2092}, \href
  {http://adsabs.harvard.edu/abs/2016MNRAS.455.1665B} {455, 1665}

\bibitem[\protect\citeauthoryear{Brown et~al.,}{Brown
  et~al.}{2014}]{BrownDragonfly}
Brown A.~J.,  et~al., 2014, in 2014 International Conference on
  Electromagnetics in Advanced Applications (ICEAA). pp 268--271,
  \mndoi{10.1109/ICEAA.2014.6903860}

\bibitem[\protect\citeauthoryear{Carozzi \& Woan}{Carozzi \&
  Woan}{2011}]{carozzi2011fundamental}
Carozzi T.,  Woan G.,  2011, \mndoi [IEEE Transactions on Antennas and
  Propagation] {10.1109/TAP.2011.2123862}, \href
  {http://adsabs.harvard.edu/abs/2011ITAP...59.2058C} {59, 2058}

\bibitem[\protect\citeauthoryear{{Champion} et~al.,}{{Champion}
  et~al.}{2016}]{champion2016five}
{Champion} D.~J.,  et~al., 2016, \mndoi [\mnras] {10.1093/mnrasl/slw069}, \href
  {http://adsabs.harvard.edu/abs/2016MNRAS.460L..30C} {460, L30}

\bibitem[\protect\citeauthoryear{Chippendale, Beresford, Deng, Leach, Reynolds,
  Kramer  \& Tzioumis}{Chippendale et~al.}{2016}]{chippendale2016testing}
Chippendale A.~P.,  Beresford R.~J.,  Deng X.,  Leach M.,  Reynolds J.~E.,
  Kramer M.,   Tzioumis T.,  2016, in 2016 International Conference on
  Electromagnetics in Advanced Applications (ICEAA). pp 909--912,
  \mndoi{10.1109/ICEAA.2016.7731550}

\bibitem[\protect\citeauthoryear{Cordes, Kramer, Lazio, Stappers, Backer  \&
  Johnston}{Cordes et~al.}{2004}]{cordes2004pulsars}
Cordes J.,  Kramer M.,  Lazio T.,  Stappers B.,  Backer D.,   Johnston S.,
  2004, \mndoi [New Astronomy Reviews] {10.1016/j.newar.2004.09.040}, \href
  {http://adsabs.harvard.edu/abs/2004NewAR..48.1413C} {48, 1413}

\bibitem[\protect\citeauthoryear{{Dai} et~al.,}{{Dai}
  et~al.}{2015}]{ShiPolProfiles}
{Dai} S.,  et~al., 2015, \mndoi [\mnras] {10.1093/mnras/stv508}, \href
  {http://adsabs.harvard.edu/abs/2015MNRAS.449.3223D} {449, 3223}

\bibitem[\protect\citeauthoryear{DeBoer et~al.,}{DeBoer
  et~al.}{2009}]{DeBoer2009}
DeBoer D.~R.,  et~al., 2009, \mndoi [Proceedings of the IEEE]
  {10.1109/JPROC.2009.2016516}, \href
  {http://adsabs.harvard.edu/abs/2009IEEEP..97.1507D} {97, 1507}

\bibitem[\protect\citeauthoryear{Dunning, Bowen, Bourne, Hayman  \&
  Smith}{Dunning et~al.}{2015}]{Dunning2015}
Dunning A.,  Bowen M.,  Bourne M.,  Hayman D.,   Smith S.~L.,  2015, in
  Antennas and Propagation in Wireless Communications (APWC), 2015 IEEE-APS
  Topical Conference on. pp 787--790, \mndoi{10.1109/APWC.2015.7300180}

\bibitem[\protect\citeauthoryear{{Edwards}, {Hobbs}  \& {Manchester}}{{Edwards}
  et~al.}{2006}]{R-Edwards_2006}
{Edwards} R.~T.,  {Hobbs} G.~B.,   {Manchester} R.~N.,  2006, \mndoi [\mnras]
  {10.1111/j.1365-2966.2006.10870.x}, \href
  {http://adsabs.harvard.edu/abs/2006MNRAS.372.1549E} {372, 1549}

\bibitem[\protect\citeauthoryear{Foster, Karastergiou, Paulin, Carozzi,
  Johnston  \& van Straten}{Foster et~al.}{2015}]{foster2015intrinsic}
Foster G.,  Karastergiou A.,  Paulin R.,  Carozzi T.,  Johnston S.,   van
  Straten W.,  2015, \mndoi [Monthly Notices of the Royal Astronomical Society]
  {10.1093/mnras/stv1722}, \href
  {http://adsabs.harvard.edu/abs/2015MNRAS.453.1489F} {453, 1489}

\bibitem[\protect\citeauthoryear{Hampson et~al.,}{Hampson
  et~al.}{2012}]{hampson2012askap}
Hampson G.,  et~al., 2012, in 2012 International Conference on Electromagnetics
  in Advanced Applications. pp 807--809, \mndoi{10.1109/ICEAA.2012.6328742}

\bibitem[\protect\citeauthoryear{Hampson, Brown, Bunton, Neuhold, Chekkala,
  Bateman  \& Tuthill}{Hampson et~al.}{2014}]{HampsonRedback}
Hampson G.~A.,  Brown A.,  Bunton J.~D.,  Neuhold S.,  Chekkala R.,  Bateman
  T.,   Tuthill J.,  2014, in 2014 XXXIth URSI General Assembly and Scientific
  Symposium (URSI GASS). pp~1--4, \mndoi{10.1109/URSIGASS.2014.6930062}

\bibitem[\protect\citeauthoryear{Hay \& O'sullivan}{Hay \&
  O'sullivan}{2008}]{Hay2008}
Hay S.,  O'sullivan J.,  2008, \mndoi [Radio Science] {10.1029/2007RS003798},
  43

\bibitem[\protect\citeauthoryear{{Hobbs}, {Edwards}  \& {Manchester}}{{Hobbs}
  et~al.}{2006}]{G-Hobbs_2006}
{Hobbs} G.~B.,  {Edwards} R.~T.,   {Manchester} R.~N.,  2006, \mndoi [\mnras]
  {10.1111/j.1365-2966.2006.10302.x}, \href
  {http://adsabs.harvard.edu/abs/2006MNRAS.369..655H} {369, 655}

\bibitem[\protect\citeauthoryear{{Hobbs} et~al.,}{{Hobbs}
  et~al.}{2011}]{hobbs2011parkes}
{Hobbs} G.,  et~al., 2011, \mndoi [\pasa] {10.1071/AS11016}, \href
  {http://adsabs.harvard.edu/abs/2011PASA...28..202H} {28, 202}

\bibitem[\protect\citeauthoryear{{Hobbs} et~al.,}{{Hobbs}
  et~al.}{2016}]{hobbs2016pilot}
{Hobbs} G.,  et~al., 2016, \mndoi [\mnras] {10.1093/mnras/stv2893}, \href
  {http://adsabs.harvard.edu/abs/2016MNRAS.456.3948H} {456, 3948}

\bibitem[\protect\citeauthoryear{{Hotan}, {van Straten}  \&
  {Manchester}}{{Hotan} et~al.}{2004}]{hvm04}
{Hotan} A.~W.,  {van Straten} W.,   {Manchester} R.~N.,  2004, \mndoi [\pasa]
  {10.1071/AS04022}, \href {http://adsabs.harvard.edu/abs/2004PASA...21..302H}
  {21, 302}

\bibitem[\protect\citeauthoryear{{Johnston}}{{Johnston}}{2002}]{JohnstonCalibration}
{Johnston} S.,  2002, \mndoi [\pasa] {10.1071/AS01062}, \href
  {http://adsabs.harvard.edu/abs/2002PASA...19..277J} {19, 277}

\bibitem[\protect\citeauthoryear{{Keane} \& {SUPERB Collaboration}}{{Keane} \&
  {SUPERB Collaboration}}{2016}]{keane2016fast}
{Keane} E.~F.,  {SUPERB Collaboration} 2016, preprint, \href
  {http://adsabs.harvard.edu/abs/2016arXiv160205165K} {} (\mn@eprint {arXiv}
  {1602.05165})

\bibitem[\protect\citeauthoryear{{Keith} et~al.,}{{Keith}
  et~al.}{2010}]{keith2010high}
{Keith} M.~J.,  et~al., 2010, \mndoi [\mnras]
  {10.1111/j.1365-2966.2010.17325.x}, \href
  {http://adsabs.harvard.edu/abs/2010MNRAS.409..619K} {409, 619}

\bibitem[\protect\citeauthoryear{{Keith}, {Shannon}  \& {Johnston}}{{Keith}
  et~al.}{2013}]{keith2013connection}
{Keith} M.~J.,  {Shannon} R.~M.,   {Johnston} S.,  2013, \mndoi [\mnras]
  {10.1093/mnras/stt660}, \href
  {http://adsabs.harvard.edu/abs/2013MNRAS.432.3080K} {432, 3080}

\bibitem[\protect\citeauthoryear{{Kerr} et~al.,}{{Kerr}
  et~al.}{2014}]{kerr2014young}
{Kerr} M.,  et~al., 2014, {Young Pulsar Timing and the Fermi Mission}, ATNF
  Proposal

\bibitem[\protect\citeauthoryear{{Lorimer} et~al.,}{{Lorimer}
  et~al.}{2015}]{lorimer2015parkes}
{Lorimer} D.~R.,  et~al., 2015, \mndoi [\mnras] {10.1093/mnras/stv804}, \href
  {http://adsabs.harvard.edu/abs/2015MNRAS.450.2185L} {450, 2185}

\bibitem[\protect\citeauthoryear{{Manchester}, {Hobbs}, {Teoh}  \&
  {Hobbs}}{{Manchester} et~al.}{2005}]{manchesteraustralia}
{Manchester} R.~N.,  {Hobbs} G.~B.,  {Teoh} A.,   {Hobbs} M.,  2005, \mndoi
  [\aj] {10.1086/428488}, \href
  {http://adsabs.harvard.edu/abs/2005AJ....129.1993M} {129, 1993}

\bibitem[\protect\citeauthoryear{Manchester, {(for the PPTA Team),}, Caretti,
  Norris  \& Phillips}{Manchester et~al.}{2013a}]{ManchesterPPTA2013}
Manchester R.~N.,  {(for the PPTA Team),} Caretti E.,  Norris R.~P.,   Phillips
  C.~J.,  2013a, ATNF Technical Memo 40.3.2/002, Development of an
  Ultra-Wideband (UWL) Receiver System at Parkes.
CSIRO
  \href{http://www.atnf.csiro.au/observers/memos/Parkes_UWL_proposal.pdf}{\nolinkurl{http://www.atnf.csiro.au/observers/memos/}}\\\href{http://www.atnf.csiro.au/observers/memos/Parkes_UWL_proposal.pdf}{\nolinkurl{Parkes_UWL_proposal.pdf}}

\bibitem[\protect\citeauthoryear{{Manchester} et~al.,}{{Manchester}
  et~al.}{2013b}]{manchester2013parkes}
{Manchester} R.~N.,  et~al., 2013b, \mndoi [\pasa] {10.1017/pasa.2012.017},
  \href {http://adsabs.harvard.edu/abs/2013PASA...30...17M} {30, e017}

\bibitem[\protect\citeauthoryear{{McConnell} et~al.,}{{McConnell}
  et~al.}{2016}]{McConnellBETA2016}
{McConnell} D.,  et~al., 2016, \mndoi [\pasa] {10.1017/pasa.2016.37}, \href
  {http://adsabs.harvard.edu/abs/2016PASA...33...42M} {33, e042}

\bibitem[\protect\citeauthoryear{{Ng} et~al.,}{{Ng} et~al.}{2015}]{ng2015high}
{Ng} C.,  et~al., 2015, \mndoi [\mnras] {10.1093/mnras/stv753}, \href
  {http://adsabs.harvard.edu/abs/2015MNRAS.450.2922N} {450, 2922}

\bibitem[\protect\citeauthoryear{{Ravi} et~al.,}{{Ravi}
  et~al.}{2016}]{ravi2016magnetic}
{Ravi} V.,  et~al., 2016, \mndoi [Science] {10.1126/science.aaf6807}, \href
  {http://adsabs.harvard.edu/abs/2016Sci...354.1249R} {354, 1249}

\bibitem[\protect\citeauthoryear{Sarkissian, Reynolds, Hobbs  \&
  Harvey-Smith}{Sarkissian et~al.}{2017}]{Sarkissian2017}
Sarkissian J.,  Reynolds J.,  Hobbs G.,   Harvey-Smith L.,  2017, Monitoring
  the Vela pulsar for 1 year using a Phased Array Feed, unpublished

\bibitem[\protect\citeauthoryear{Sault}{Sault}{2014}]{SaultBETA2014}
Sault R.~J.,  2014, ACES MEMO~2, Initial characterisation of BETA polarimetric
  response.
CSIRO
  \href{http://www.atnf.csiro.au/projects/askap/ACES-memos}{\nolinkurl{http://www.atnf.csiro.au/projects/askap/}}\\\href{http://www.atnf.csiro.au/projects/askap/ACES-memos}{\nolinkurl{ACES-memos}}

\bibitem[\protect\citeauthoryear{Schinckel \& Bock}{Schinckel \&
  Bock}{2016}]{SchinckelASKAP2016}
Schinckel A. E.~T.,  Bock D. C.-J.,  2016, in Proc. SPIE. pp 99062A--99062A--9,
  \mndoi{10.1117/12.2233920}

\bibitem[\protect\citeauthoryear{{Shannon} et~al.,}{{Shannon}
  et~al.}{2014}]{shannon2014limitations}
{Shannon} R.~M.,  et~al., 2014, \mndoi [\mnras] {10.1093/mnras/stu1213}, \href
  {http://adsabs.harvard.edu/abs/2014MNRAS.443.1463S} {443, 1463}

\bibitem[\protect\citeauthoryear{{Staveley-Smith} et~al.,}{{Staveley-Smith}
  et~al.}{1996}]{staveley1996parkes}
{Staveley-Smith} L.,  et~al., 1996, \mndoi [\pasa]
  {https://doi.org/10.1017/S1323358000020919}, \href
  {http://adsabs.harvard.edu/abs/1996PASA...13..243S} {13, 243}

\bibitem[\protect\citeauthoryear{{Thornton} et~al.,}{{Thornton}
  et~al.}{2013}]{thornton2013population}
{Thornton} D.,  et~al., 2013, \mndoi [Science] {10.1126/science.1236789}, \href
  {http://adsabs.harvard.edu/abs/2013Sci...341...53T} {341, 53}

\bibitem[\protect\citeauthoryear{{Tiburzi} et~al.,}{{Tiburzi}
  et~al.}{2016}]{tiburzi2016study}
{Tiburzi} C.,  et~al., 2016, \mndoi [\mnras] {10.1093/mnras/stv2143}, \href
  {http://adsabs.harvard.edu/abs/2016MNRAS.455.4339T} {455, 4339}

\bibitem[\protect\citeauthoryear{Tuthill, Hampson, Bunton, Brown, Neuhold,
  Bateman, de Souza  \& Joseph}{Tuthill et~al.}{2012}]{TuthillFilterbank}
Tuthill J.,  Hampson G.,  Bunton J.,  Brown A.,  Neuhold S.,  Bateman T.,  de
  Souza L.,   Joseph J.,  2012, in 2012 International Conference on
  Electromagnetics in Advanced Applications. pp 1067--1070,
  \mndoi{10.1109/ICEAA.2012.6328788}

\bibitem[\protect\citeauthoryear{Tuthill, Hampson, Bunton, Harris, Brown,
  Ferris  \& Bateman}{Tuthill et~al.}{2015}]{Tuthill2015}
Tuthill J.,  Hampson G.,  Bunton J.,  Harris F.,  Brown A.,  Ferris R.,
  Bateman T.,  2015, in Signal Processing and Signal Processing Education
  Workshop (SP/SPE), 2015 IEEE. pp 255--260,
  \mndoi{10.1109/DSP-SPE.2015.7369562}

\bibitem[\protect\citeauthoryear{{Verbiest} et~al.,}{{Verbiest}
  et~al.}{2016}]{verbiest2016international}
{Verbiest} J.~P.~W.,  et~al., 2016, \mndoi [\mnras] {10.1093/mnras/stw347},
  \href {http://adsabs.harvard.edu/abs/2016MNRAS.458.1267V} {458, 1267}

\bibitem[\protect\citeauthoryear{Warnick, Ivashina, Wijnholds  \&
  Maaskant}{Warnick et~al.}{2012}]{warnick2012polarimetry}
Warnick K.~F.,  Ivashina M.~V.,  Wijnholds S.~J.,   Maaskant R.,  2012, \mndoi
  [{IEEE} Trans. Antennas Propag.] {10.1109/TAP.2011.2167926}, 60, 184

\bibitem[\protect\citeauthoryear{Weltevrede \& Johnston}{Weltevrede \&
  Johnston}{2008}]{weltevrede2008profile}
Weltevrede P.,  Johnston S.,  2008, \mndoi [Monthly Notices of the Royal
  Astronomical Society] {10.1111/j.1365-2966.2008.13950.x}, \href
  {http://adsabs.harvard.edu/abs/2008MNRAS.391.1210W} {391, 1210}

\bibitem[\protect\citeauthoryear{van Cappellen, Bakker  \& Oosterloo}{van
  Cappellen et~al.}{2011}]{van2011experimental}
van Cappellen W.~A.,  Bakker L.,   Oosterloo T.~A.,  2011, in 2011 XXXth URSI
  General Assembly and Scientific Symposium. pp~1--4,
  \mndoi{10.1109/URSIGASS.2011.6050405}

\bibitem[\protect\citeauthoryear{{van Straten}}{{van
  Straten}}{2004}]{van2004radio}
{van Straten} W.,  2004, \mndoi [\apjs] {10.1086/383187}, \href
  {http://adsabs.harvard.edu/abs/2004ApJS..152..129V} {152, 129}

\bibitem[\protect\citeauthoryear{{van Straten} \& {Bailes}}{{van Straten} \&
  {Bailes}}{2011}]{vanSratenDSPSR}
{van Straten} W.,  {Bailes} M.,  2011, \mndoi [\pasa] {10.1071/AS10021}, \href
  {http://adsabs.harvard.edu/abs/2011PASA...28....1V} {28, 1}

\makeatother
\end{thebibliography}
\end{document}